# ON THE INFLUENCE OF SUPERNOVA REMNANT THERMAL ENERGY IN POWERING GALACTIC WINDS

by Brad K. Gibson[1,2]

[1]*Department of Astrophysics, University of Oxford*

[2]*Department of Geophysics & Astronomy, University of British Columbia*

## Abstract

The fundamental tenet of the classical supernovae-driven wind model of elliptical galaxies is that the residual thermal energy of all supernovae remnants (SNRs) provide sufficient energy to overcome the binding energy of the remaining interstellar gas, thereby driving a global galactic wind. We re-examine model predictions of this epoch of wind ejection $t_{\rm GW}$, highlighting a heretofore underappreciated sensitivity to the adopted remnant thermal energy formalism, and illustrating cases in which previous work may have substantially overestimated $t_{\rm GW}$. Arguments based upon chemical evolution *alone*, put forth to reject the hypothesis of dark matter distributions similar to the luminous component in spheroids, are shown to be tenuous. Finally, the predicted enrichment of intracluster gas during the wind phase of cluster ellipticals, and its relation to the selected SNR interior thermal energy evolutionary scheme, is addressed. Despite the success of previous wind models, our results still call into question the correctness of the simple analytical approach used thus far, and imply that a more appropriate technique should be adopted in the future.

*1. Introduction.* It has long been recognised that the colours of elliptical galaxies become increasingly red with decreasing magnitudes, or conversely, increasing mass. This effect has been commonly attributed to an increasing average galactic metal content with increasing galactic mass (e.g. Faber 1977, for an early review). The key to understanding the origin of these correlations was provided by Mathews & Baker (1971), but not fully appreciated until Larson (1974).

In order to explain the paucity of interstellar gas in most ellipticals Mathews & Baker (1971) postulated that much of it had been strongly heated by supernovae (SNe) explosions and driven out by a hot galactic wind once the residual thermal energy of all SNe remnants exceeded the binding energy of the remaining gas, thereby bringing to a halt the bulk of active star formation. Subsequent evolution would then be regulated by the gas returned to the interstellar medium (ISM) from dying stars. It was left to Larson (1974) to recognise that the binding energy per unit mass of gas is higher in the more massive galaxies, and thus these systems would retain their gas for longer initial periods before reaching this epoch of galactic wind $t_{\rm GW}$, thereby attaining higher metallicities in a manner consistent with the observed mass-metallicity relationship. Further developments over the past decade can be attributed to Saito (1979), Arimoto & Yoshii (1987), Matteucci & Tornambé (1987), Angeletti & Giannone (1990), Bressan, Chiosi & Fagotto (1994), Elbaz, Arnaud & Vangioni-Flam (1994), and Gibson (1995). It is generally accepted that wind models of this nature may be inadequate in some details, but are still the framework of choice for modeling the early evolution of spheroids (Renzini et al. 1993).

Determining the epoch of global gas ejection is an imperative aspect of these wind models as $t_{\rm GW}$ governs the cessation of the bulk of star formation, as well as the amount and metallicity of expelled gas, and the current spectrophotometric properties of ellipticals. Angeletti & Giannone (1991) have discussed the sensitivity of $t_{\rm GW}$ predictions to the assumed initial proto-galactic radius (contributing to the binding energy of the spheroid), and Matteucci (1994) has explored recently the importance of the astration parameter $\nu$ (contributing to the timescale for star formation). Our plan herein is to re-examine briefly one of the fundamental ingredients to all supernovae-powered wind models, specifically, the formalism adopted to represent the evolution of a supernova remnant's (SNR) internal thermal energy $\varepsilon_{\rm th}(t)$, and the availability of such energy for driving a galactic wind.

In Section 2 we outline the basic chemical evolution and wind model framework used in our study. We present the classic Cox (1972) and Chevalier (1974) formalism used in each of the post-Larson (1974) papers listed above, highlighting an oversight which has lead, in some cases, to an overestimation of the gas ejection time $t_{\rm GW}$. Following Cioffi, McKee & Bertschinger (1988), we discuss their alternative formalism for the evolution of the SNR thermal energy $\varepsilon_{\rm th}(t)$, superseding the earlier Cox and Chevalier work with its more sophisticated treatment of radiative cooling in the hot SNR interior. Sections 3 and 4 provide the relevant equations governing the evolution of $\varepsilon_{\rm th}(t)$ set by these competing formalisms. A direct comparison of the differences in some of the model predictions based upon the selection of thermal energy formalism is given in Section 5 and summarised in Section 6. Our aim is not to analyse each and every consequence of this selection, but only to highlight a few important ones, and in particular, to demonstrate the influence of $\varepsilon_{\rm th}(t)$ upon $t_{\rm GW}$ predictions. This is sufficient for our purposes as many of the predictions (e.g. mass and metallicity of gas ejected) are related in a simple manner to $t_{\rm GW}$.

*2. The Classical Wind Model of Ellipticals.* For convenience we have adopted the classical one-zone wind model of ellipticals as presented by Matteucci & Greggio (1986) and Matteucci & Tornambé (1987). A brief outline follows, with a more detailed accounting of the relevant equations and input ingredients pertaining to our package (entitled **MEGaW**≡**M**etallicity **E**volution with **Ga**lactic **W**inds) to be presented elsewhere (Gibson 1995), although the reader is directed to the above references, as well as those of Arimoto & Yoshii (1987) and Angeletti & Giannone (1990), for a complementary introduction to the subject.

The gas mass $M_{\rm g}$ in the ISM is depleted by star formation and replenished through ejection from stars, and is governed by

$$\frac{{\rm d}M_{\rm g}(t)}{{\rm d}t} = -\psi(t) + E(t), \tag{1}$$

where $\psi(t)$ is the star formation rate and $E(t)$ is the total gas ejection rate from all stars. Coupled to the gas mass equation 1, the mass of metals $M_Z \equiv ZM_{\rm g}$ in the ISM evolves as

$$\frac{{\rm d}M_Z(t)}{{\rm d}t} = -Z(t)\psi(t) + E_Z(t), \tag{2}$$

where $E_Z(t)$ is the total ejection rate of both processed and unprocessed metals. The relations for $E(t)$ and $E_Z(t)$ are adapted from Greggio & Renzini (1983) and Matteucci & Greggio (1986).

The nucleosynthesis prescriptions utilised are a function of both mass and metallicity. For the purpose of this paper we have used the yields of Arnett (1991) for Type II SNe, Renzini & Voli (1981) for single low and intermediate mass stars, and Thielemann, Nomoto & Yokoi (1986) for Type Ia SNe. Details related to the yields and the assumed SNe progenitors used in **MEGaW** can be found in Gibson (1995).

The star formation rate $\psi(t)$ is taken to depend upon the gas mass $M_{\rm g}(t)$ such that

$$\psi(t) = \nu M_{\rm g}(t) \quad [{\rm M}_\odot/{\rm yr}], \tag{3}$$

where the astration parameter $\nu$ is the inverse of the star formation timescale, which for a galaxy of mass $M_{\rm G}$ is given by (Arimoto & Yoshii 1987)

$$\nu = 8.6(M_{\rm G}/10^{12}M_\odot)^{-0.115} \quad [{\rm Gyr}^{-1}]. \tag{4}$$

Recall that for $t > t_{\rm GW}$, the star formation rate $\psi(t)$ is taken to be zero.

We use a singular power law of slope $x = 1.35$ (Salpeter 1955) or $x = 0.95$ (Arimoto & Yoshii 1987) for the IMF, normalised to unity over the mass range $m_\ell = 0.1$ M$_\odot$ to $m_u = 120.0$ M$_\odot$. Güsten & Mezger (1983) stellar lifetimes are used, and their respective remnant masses are taken from Maeder (1992).

As alluded to previously, the time at which global gas ejection occurs in the classical wind model $t_{\rm GW}$ is set by the residual thermal energy of the interstellar gas $E_{\rm th}$ overcoming the binding energy of said gas $\Omega_{\rm g}$ (Larson 1974), where

$$E_{\rm th}(t) = \int_0^t \varepsilon_{\rm th}(t-t')R_{\rm SN}(t'){\rm d}t'. \tag{5}$$

$t'$ is the SN explosion time. $R_{\rm SN}(t')$ is the total SNe rate (Type Ia and II) at time $t'$, and is taken taken from Greggio & Renzini (1983) and Matteucci & Tornambé (1987). Finally, $\varepsilon_{\rm th}$ represents the evolution of the thermal energy content in the interior of a SNR.

$\Omega_{\rm g}(t)$ is sensitive to the assumed dark matter distribution. Matteucci (1992) finds that for realistic diffuse halo distributions, the effect upon $t_{\rm GW}$, and the chemical evolution in general, is negligible. Hence we shall defer discussion of the diffuse dark matter halo model to her paper, as well as to recent work by Bressan, Chiosi & Fagotto (1994) and Forman, Jones & Tucker (1994). For our purposes, we shall use the binding energy formalism appropriate for a two-component spheroid (dark matter + luminous matter) in which the components are distributed similarly, and the ratio of the masses of dark to luminous matter is denoted $R$. Thus, using Saito (1979) and Matteucci (1992), we have

$$\Omega_{\rm g}(t) = \Omega_{\rm G}\frac{M_{\rm g}(t)}{M_{\rm g}(0)}\left(2.0 - \frac{M_{\rm g}(t)}{M_{\rm g}(0)}\right), \tag{6}$$

where the initial binding energy of the system $\Omega_{\rm G}$ of initial mass $M_{\rm G} \equiv M_{\rm lum}(1+R)$ (in grams) is

$$\Omega_{\rm G} = 3.41 \times 10^{-6} M_{\rm G}^{1.45} \quad [{\rm erg}]. \tag{7}$$

We can see from equations 6 and 7 that the ratio of dark to luminous mass $R$ has the effect of increasing the binding energy of the gas by a factor of $(1+R)^{1.45}$ and the virial radius of the spheroid by $(1+R)^{0.55}$ (e.g. equation (6) of Matteucci 1992). What this means for predicted epochs of gas ejection will be addressed in Section 5.

Finally, the exact formalism adopted for $\varepsilon_{\rm th}$ impacts greatly upon the predicted values of $t_{\rm GW}$, and hence upon the predictions of mass of metals ejected at $t_{\rm GW}$, as we shall discuss in Section 5. First though we shall introduce the competing formalisms available describing the evolution of $\varepsilon_{\rm th}$, including the classic formulation of Cox (1972) and Chevalier (1974) (hereafter referred to as the "CC formalism"), as well as the newer work of Cioffi, McKee & Bertschinger (1988) (hereafter, the "CMB formalism"). It is important to stress that all of the previous wind models listed earlier (e.g. Saito 1979; Arimoto & Yoshii 1987; Matteucci & Tornambé 1987; Angeletti & Giannone 1990; Bressan, Chiosi & Fagotto 1994; Elbaz, Arnaud & Vangioni-Flam 1994; Gibson 1994) use the CC formalism. The work presented herein is the first to explore usage of the CMB form.

*3. Supernova Remnant Thermal Energy Evolution: CMB Formalism.* The evolution of a SNR can be characterised by three dynamical phases: (i) free expansion until the mass of interstellar material swept up reaches that of the SN ejecta, at which time a shock wave forms; (ii) adiabatic expansion which continues until the radiative cooling time of newly shocked gas equals the expansion time of the remnant; and, (iii) formation of a cold dense shell behind the front which begins when some sections of the shocked gas have radiated most of their thermal energy, being further compressed by the pressure of the remainder of the shocked material. Comprehensive reviews of the relevant shock dynamics include Ostriker & McKee (1988) and Draine & McKee (1993).

Using a 1d-Lagrangian hydrodynamics code, Cioffi, McKee & Bertschinger (1988) (CMB) followed the dynamical and thermal evolution of SNRs from the initial ejecta phase, into the adiabatic phase, and through to the phase in which radiative cooling of the hot interior becomes important. By insisting that the expansion of the shock radius $R_{\rm s}$ obey a power law in time (i.e. $R_{\rm s} \propto t^{\eta}$), CMB identified each of the phases simply by adopting the appropriate slope $\eta$ for each stage, the value being governed by a series of fits to their numerical models.

The initial free expansion phase enters the adiabatic phase once the swept-up interstellar material begins to dominate the remnant mass, at which point the shock front slows down. The familiar Sedov-Taylor solution for a self-similar adiabatic shock with $\eta_{\rm ST} = 2/5$ is encountered ($t$ in s):

$$R_{\rm s} = \left(\frac{\xi E_0}{\rho_0}\right)^{1/5} t^{2/5} \quad [\mathrm{cm}], \tag{8}$$

where $R_{\rm s}$ is the radius of the outer edge of the SNR shock front, $E_0 \equiv \varepsilon_0 \times 10^{51}$ erg is the initial blast energy (i.e. $\varepsilon_0$ represents said energy in units of $10^{51}$ erg), $\rho_0$ is the ambient mass density (g/cm$^3$), and the constant $\xi = 2.026$ (appropriate for $\gamma = 5/3$) (Ostriker & McKee 1988).

Radiative cooling of the shocked material leads to the formation of a thin, dense shell at time $t_{\rm sf}$, where

$$t_{\rm sf} = 3.61 \times 10^4 \varepsilon_0^{3/14} \zeta_m^{-5/14} n_0^{-4/7} \quad [\mathrm{yr}], \tag{9}$$

which "snowplows" through the ISM, driven from behind by the pressure of the hot SNR interior, as well as its own momentum. $\zeta_m$ is a parameter which takes into account the influence of metallicity in determining the shell cooling time ($\zeta_m \equiv Z/Z_\odot$). $n_0$ is the Hydrogen number density (cm$^{-3}$) in the surrounding ambient gas. A recent study of thin shell formation timescales corroborates the use of equation 9 for typical uniform ambient ISM densities (Franco et al. 1994).

Before $t_{\rm sf}$, at time $t_{\rm pds} \approx 0.37 t_{\rm sf}$, radiative losses begin to affect the evolution of the remnant. At this point, the post-shock fluid velocity approaches the shock velocity, entering the the so-called "pressure-driven snowplow" (PDS) phase. If one ignores cooling of the hot interior, the dynamics of the shell are governed by $R_{\rm s} \propto t^{2/7}$ (Ostriker & McKee 1988). CMB found a more suitable fit in the PDS stage using $\eta_{\rm pds} = 3/10$, as opposed to this standard value of 2/7. Specifically, they recommend the PDS solution

$$R_{\rm s} = R_{\rm pds}\left(\frac{4t}{3t_{\rm pds}} - \frac{1}{3}\right)^{3/10}, \tag{10}$$

where the radius at the beginning of the PDS stage is

$$R_{\rm pds} = 14.0\varepsilon_0^{2/7} n_0^{-3/7} \zeta_m^{-1/7} \quad [{\rm pc}]. \tag{11}$$

Recall that $t_{\rm pds}$ in equation 10 is given by $t_{\rm pds} \approx 0.37 t_{\rm sf}$.

Fitting to their simulations, CMB then present the following empirical representation for the evolution of the interior thermal energy of a SNR $\varepsilon_{\rm th}(t)$ for times $t < a_2 t_{\rm sf}$, where the constant $a_2 = 1.169$:

$$\varepsilon_{\rm th}(t) = 0.717 E_0 a_1 \left[1 - \left(\frac{t}{a_2 t_{\rm sf}}\right)^{14/5}\right] + \frac{0.717 E_0 (1 - a_1)}{\left[\left(\frac{R_{\rm s}}{R_{\rm sf}}\right)^{10} + 1\right]^{1/5} \left[\left(\frac{t}{t_{\rm sf}}\right)^{4} + 1\right]^{1/9}}, \tag{12}$$

where $a_1 = 0.398$. Thus, the thermal energy evolution during the early dynamical phases of a SNR's history can be determined by coupling equations 8 ($0 \le t \le t_{\rm PDS}$), 9, and 10 ($t_{\rm PDS} \le t \le a_2 t_{\rm sf}$), with equation 12.

The interior gas continues to lose energy by pushing the shell through the ISM ($\varepsilon_{\rm th} \propto t^{-3/5}$), but there is an additional source of energy loss in the PDS phase, that being radiative cooling ($\varepsilon_{\rm th} \propto t^{-4/9}$). CMB provide the first analytical treatment of SNR evolution which incorporates explicitly these radiative cooling losses from the interior. The late-time behaviour of their remnants does approach the expected "momentum-conserving snowplow" (MCS) solution of Oort (1951), with $R_{\rm s} \propto t^{1/4}$ (at time $t \gtrsim t_{\rm mcs}$), eventually merging with the ISM when the internal driving pressure vanishes due to the radiative cooling in the interior, *provided* that this radiative cooling has not robbed the interior of the necessary hot gas to continue driving the shell through the MCS stage. In general though, CMB find that their remnants merge with the ISM before ever entering the MCS phase because the interior pressure drops below the ambient ISM pressure at times much less than $t_{\rm mcs}$.

Similar to equation 12, CMB present an empirical form for the SNR interior thermal energy evolution $\varepsilon_{\rm th}(t)$ during this later PDS phase, valid for times $t \ge a_2 t_{\rm sf}$:

$$\varepsilon_{\rm th}(t) = \frac{0.717 E_0 (1 - a_1)}{\left[\left(\frac{R_{\rm s}}{R_{\rm sf}}\right)^{10} + 1\right]^{1/5} \left[\left(\frac{t}{t_{\rm sf}}\right)^{4} + 1\right]^{1/9}}. \tag{13}$$

In the CMB formalism, expansion of the remnant is halted at a radius $R_{\rm merge}$ once the interior pressure in the SNR is reduced to that of the pressure of the ambient ISM, thereafter merging and becoming indistinguishable from the surrounding ISM, at time $t_{\rm merge}$, where

$$t_{\rm merge} = 57 t_{\rm pds} \varepsilon_0^{5/49} n_0^{10/49} \zeta_m^{15/49} \tag{14}$$

and

$$R_{\rm merge} = 3.7 R_{\rm pds} \varepsilon_0^{3/98} n_0^{3/49} \zeta_m^{9/98}. \quad (15)$$

Beyond $t_{\rm merge}$, the thermal evolution is no longer governed by the combination of the expansion term ($\varepsilon_{\rm th} \propto t^{-3/5}$) and the radiative cooling term ($\varepsilon_{\rm th} \propto t^{-4/9}$) in equation 13, but by only the latter. This late-time scenario, denoted Model $B_2$ in our work, is computed by fixing $R_{\rm s}$ in equation 13 to be $R_{\rm merge}$ (equation 15), for all times $t > t_{\rm merge}$. Typically though, the radiative cooling only proceeds for a time $t_{\rm cool}$, where

$$t_{\rm cool} \approx 550 t_{\rm pds} \zeta_m^{-9/14}, \quad (16)$$

until the hot interior gas has cooled to $T \sim 10^4$ K, below which radiative losses become minimal (e.g. Sutherland & Dopita 1993; Saito 1979). Our Model $B_3$ follows this prescription, depositing the residual SNR interior thermal energy at time $t_{\rm cool}$ into the ISM with no subsequent energy loss.

While Model $B_3$ is our preferred conservative model, alternative late time thermal energy evolution scenarios have been explored. Model $B_1$ is similar to the model used by Larson (1974), in which radiative losses are negligible after the remnant has merged with the ISM (at time $t \equiv t_{\rm merge}$). Model $B_0$ is the opposite extreme; in this case, SNRs continue to expand in the PDS phase and cool radiatively for all times. As will be shown in Section 5, Model $B_0$ leads to unfeasible results.

Figure 1 highlights the differences in the late-time thermal evolution $\varepsilon_{\rm th}$ for the four different models explored using the CMB formalism. The example shown represents a SN blast of initial energy $E_0 = 10^{51}$ erg, expanding into an ambient homogeneous ISM of number density $n_0 = 0.55\,{\rm cm}^{-3}$, mass density $\rho_0 = 9.2\times10^{-25}$ g/cm$^3$, and solar metallicity. Model $B_0$ is taken to expand and cool radiatively for all times (i.e. $\varepsilon_{\rm th}(t) \propto t^{-1.04}$). The remaining B models trace the evolution of Model $B_0$ up until $t \equiv t_{\rm merge}$, the time at which pressure equilibrium between the interior and exterior ISM is reached, and expansion is halted. If radiative cooling can be neglected during the subsequent evolution, then $\varepsilon_{\rm th}(t) \equiv \varepsilon_{\rm th}(t_{\rm merge})$, as was done for Model $B_1$. In the more realistic scenario, radiative cooling reduces the SNR interior thermal energy such that $\varepsilon_{\rm th}(t) \propto t^{-0.44}$, until the time $t_{\rm cool}$ at which point most of the hot gas within the remnant has cooled to $T \sim 10^4$ K, at which point further decreases in the residual thermal energy are halted, as in Model $B_3$. Model $B_2$ continues to evolve like $\varepsilon_{\rm th} \propto t^{-0.44}$ for all times $t > t_{\rm merge}$.

Models $B_1$, $B_2$, and $B_3$ all stop expanding at time $t_{\rm merge}$, once the internal pressure in the SNR is balanced by the ambient external pressure in the ISM. Following Larson (1974), we have also computed a parallel set of models (Models $B_1'$, $B_2'$, and $B_3'$) which end their respective PDS phases at $t'_{\rm merge}$, where

$$t'_{\rm merge} \approx 5.7 \times 10^5 \varepsilon_0^{-0.34} n_0^{0.38} \psi^{-0.52} R_{\rm G}^{1.56}. \quad (17)$$

$t'_{\rm merge}$ represents the characteristic time at which the SNRs collide and merge with other SNRs, thus halting further expansion. The star formation rate $\psi$ and galactic radius $R_{\rm G}$ are both set by the chemical evolution code **MEGaW** (Gibson 1995). The differences in adopting equation 17 as opposed to equation 14 as the end of the PDS phase are usually minimal (see Section 5), but for completeness we do include this option.

The dynamical ($R_s \propto t^\eta$) and thermal ($\varepsilon_{\rm th} \propto t^\nu$) characteristics of the various models introduced above are summarised in Table I.

*4. Supernova Remnant Thermal Energy Evolution: CC Formalism.* In the pioneering work of Cox (1972) and Chevalier (1974) (together referred to as CC, henceforth), as for the CMB formalism, the dynamics of the early stages of expansion are represented by the Sedov-Taylor similarity solution for an adiabatic blast wave (equation 8).

In contrast with equation 9, the complete radiative cooling time of the SNR shell $t_c$ is given by

$$t_c = 5.7 \times 10^4 \varepsilon_0^{4/17} n_0^{-9/17} \quad [\mathrm{yr}]. \tag{18}$$

$t_c$ from equation 18 is typically a factor of two larger than $t_{sf}$ as given by equation 9 in the CMB formalism, a difference found to be negligible in predicting the epoch of wind ejection $t_{\rm GW}$.

The thermal energy in the interior of a SNR during the adiabatic expansion phase ($t < t_c$) is simply given by

$$\varepsilon_{\rm th}(t) = 0.72 E_0, \tag{19}$$

where, as before, the initial blast energy $E_0 \equiv \varepsilon_0 \times 10^{51}$ erg.

Unlike that for the CMB formalism, CC ignore the transition from the Sedov-Taylor solution to the pressure-driven snowplow, which following the notation of CMB occurs at $t_{\rm pds} < t < t_{\rm sf}$ (see Figure 1). Hence, $t_{\rm pds}$ and $t_c$ coincide in the CC formalism, and a discontinuity is encountered at the entrance to the PDS phase.

The dynamics of the SNR shell during the subsequent PDS phase follows an $R_s \propto t^{0.31}$ law (Chevalier 1974). Neither Cox (1972) nor Chevalier (1974) have included radiative cooling of the hot interior, assuming that the cooling time is always longer than the dynamical time. In their formalism, the PDS ($t > t_c$) SNR interior thermal energy evolves as

$$\varepsilon_{\rm th}(t) = 0.22 E_0 (t/t_c)^{-0.62}. \tag{20}$$

The thermal energy in the CMB formalism cools more rapidly ($\varepsilon_{\rm th} \propto t^{-1.04}$, as opposed to $\varepsilon_{\rm th} \propto t^{-0.62}$) due to this inclusion of radiative cooling. Again, see Figure 1 for an example.

As discussed in the previous subsection, we have chosen to test two different late-time thermal evolution histories. First, we allow all remnants to expand in the PDS phase for all times $t > t_c$. This scenario is denoted Model $\mathrm{A}_0$ and is the procedure adopted by all previous researchers, including Arimoto & Yoshii (1987), Matteucci & Tornambé (1987), Angeletti & Giannone (1990), Bressan, Chiosi & Fagotto (1994), Elbaz, Arnaud & Vangioni-Flam (1994), and Gibson (1994). Again, a more appropriate approach is to halt the expansion at $t_{\rm merge}$ (Model $\mathrm{A}_1$) or $t'_{\rm merge}$ (Model $\mathrm{A}'_1$ and Larson (1974)), as the SNRs will *not* expand freely for all times; they will either come into pressure equilibrium with the external ISM, or start coming into contact with neighbouring shells.

The power law behaviour of the dynamics of the shells and interior thermal energy for the models adopted under the CC formalism are listed in Table II.

*5. Discussion.* In column 2 of Table III we list the epoch of global galactic winds $t_{\rm GW}$ for three different initial galactic masses $M_{\rm g}(0)$ based upon the assumption of the models for late-time evolution of the SNR interior thermal energy $\varepsilon_{\rm th}(t)$ described in Sections 3 and 4. Recall that previous work (e.g. Arimoto & Yoshii 1987; Matteucci & Tornambé 1987; Angeletti & Giannone 1990; Bressan, Chiosi & Fagotto 1994; Elbaz, Arnaud & Vangioni-Flam 1994; Gibson 1994) adopted the CC formalism Model $A_0$. Columns 3, 4, and 5 show the gas fraction ejected at $t_{\rm GW}$, the metallicity of said gas, and hence, the mass of metals ejected from the galaxy. The tabulated results are for spheroids without a dark matter component (i.e. $R = 0$ in equation 7).

A comparison of the predictions of Models $A_0$ and $A_1$ proves particularly enlightening, as they provide an extreme example of the sensitivity to $\varepsilon_{\rm th}(t)$. Both follow the CC formalism used in the above papers (i.e. $\varepsilon_{\rm th}(t > t_{\rm c}) \propto t^{-0.62}$), but whereas Model $A_0$ assumes that all SNRs expand and cool in the PDS phase *ad infinitum*, this expansion (and hence cooling) is constrained in Model $A_1$ to halt at time $t_{\rm merge}$ once the shells come into pressure equlibrium with the ambient ISM. Alternatively, the expansion can be taken to halt once the SNRs start to overlap at time $t'_{\rm merge}$ (Larson 1974). As can be seen in Table III, this scenario (Model $A'_1$) leads to ejection timescales of the same order of Model $A_1$, and thus for brevity, will be neglected henceforth. Regardless of initial mass, the epoch of gas ejection for Model $A_0$ is five to six times later than that for Model $A_1$. For massive galaxies this implies that ~20 times the amount of gas and metals should be ejected at $t_{\rm GW}$ in the Model $A_1$ scenario. For the lower mass $A_1$ models, the ejection time occurs before chemical evolution has enriched the gas beyond about solar metallicity. Thus the increase in metal ejection is simply due to the greatly increased mass of gas ejected. For the higher mass model, the chemical enrichment has reached approximately the same level as Model $A_0$.

As might be expected from inspection of Figure 1, Model $A_0$ is quantitatively similar to that of the CMB Model $B_2$. As well, Model $B_1$ parallels Model $A_1$ in that expansion of the SNR shell halts at time $t_{\rm merge}$ (pressure equilibrium with the ambient ISM), although the resultant wind epochs occur somewhat later because the thermal energy content during the PDS phase is lower at a given time for the CMB form.

One interesting result of using the CMB formalism is seen in the $B_0$ models. These are similar to $A_0$ models in that the SNR shells expand like $R_{\rm s} \propto t^{0.3}$ (i.e. $\varepsilon_{\rm th} \propto R_{\rm s}^{-2}$) *ad infinitum*. The key difference is that CMB also include radiative cooling of the hot SNR interior, the cooling term going as $t^{-0.44}$. The influence this additional cooling has for the residual thermal energy available to heat the ISM gas at a given time is substantial. For masses $\gtrsim 10^{10}$ $M_\odot$, no galactic wind ever forms due to the increased cooling of the SNR interior. This fact alone shows that a direct replacement of the older CC form for Model $A_0$ with the newer CMB Model $B_0$ is not valid.

It is important to stress that all the models discussed here are at best a crude representation of the reality of ISM gas thermal evolution. In all the galactic wind papers discussed thus far, the ISM is assumed to be single smooth phase, whereas in reality the situation is much more complex. Three phase models (e.g. Ostriker & McKee 1988: cold cloud nuclei, surrounded by warm neutral gas, embedded in a hot thin corona) with substantial inhomogeneities are a step in the right direction, but difficult to quantify in a simple analytical form for use in galactic wind modeling (C.F. McKee 1994, priv comm). Hydrodynamical simulations of galactic winds from ellipticals, with (e.g. Missoulis 1994) and without (e.g. Ishimaru et al. 1994; Friaça & Terlevich 1994) cloudy multi-phase ISMs, are only now becoming feasible. A recent analytical multi-phase wind model has been proposed by Ferrini & Poggianti (1993), albeit at the expense of a large number of free parameters.

Of even greater importance than the imposed single phase elliptical ISM may be the influence of magnetic field and cosmic ray energies in setting the timescale for gas ejection. It seems apparent that at least for the case of the local galactic disk (e.g. Duric 1988, and references therein), the magnitude of cosmic ray, thermal gas, and interstellar magnetic field energy densities are comparable to one another. It has been demonstrated that cosmic rays travelling through a magnetised plasma with mean velocity greater than the Alfvén speed can become coupled to the plasma by the emission of magnetohydrodynamic waves, thereby "dragging" along the thermal gas as the rays escape the galaxy (Ipavich 1975). For galaxies of mass $\sim 10^{10}$ $M_\odot$ and magnetic field strengths similar to that found locally in our Galaxy (i.e. $B \sim 5\mu$G), Ipavich (1975), Zank (1989), and Fichtner et al. (1991), for example, have demonstrated that cosmic ray-driven galactic mass loss rates of 1-10 $M_\odot$/yr can be generated. Even if such winds are sustained for only 0.1-1 Gyr, the ejected mass of gas and metals would be substantial.

Magnetic fields in the ISM also play another, secondary, role in SNe-driven wind models, by influencing the three dimensional dynamical evolution of an individual SN remnant. If the magnetic field strength is high enough for the outgoing shock to be weak, then most of the initial explosion energy $E_0$ is used up doing work against the field and the initial expansion is halted rapidly, and no adiabatic Sedov-Taylor phase is encountered. Further expansion is then only possible along the direction of the interstellar field, with the late-time evolution being characterised by contraction of its equatorial regions and stretching in the axial direction (Insertis & Rees 1991). This "strong" field (e.g. $B \sim 1$G) scenario is expected to be relevant for SNe situated in the centers of galaxies, and more importantly, for those in active galactic nuclei. When the interstellar Alfvén speed is small compared to the velocity of the SN ejecta (i.e. the so-called "weak" field scenario), the initial phases are similar (i.e. the free expansion and Sedov-Taylor regimes) to the unmagnetised case. The late-time evolution is still governed by the halting of the equatorial expansion and unimpeded expansion of the shock along the field lines, again resulting in "barrel-shaped" remnants aligned with the magnetic field (e.g. Ferrière & Zweibel 1991; Tomisaka 1992).

Virtually all of the work just discussed has been tailored to studies of the Milky Way and other local spirals, simply because the dynamo-action necessary to generate large-scale galactic magnetic fields (Parker 1971) has only been confirmed in these sites. Recently though, depolarisation measurements of strong radio lobes within elliptical hosts (Garrington 1990) have implied magnetic field strengths of $\sim 10\mu$G, and the origin and amplification of these fields within the framework of the turbulent mean-field dynamo theory demonstrated (Lesch & Bender 1990), although whether the properties of these confined fields can be extrapolated and assumed to permeate the ISM of "normal" ellipticals is still a matter of debate.

The CC formalism outlined in Section 4 approximates the effects of weak (i.e. $\sim 3\mu$G) magnetic fields by adding a term to the fluid equations (Chevalier 1974). In Chevalier's analysis, the magnetic field was never energetically important unless the post-shock magnetic pressure exceeded the thermal pressure so that the magnetic field impeded shock compression. CMB neglect magnetic fields altogether but claim that they are important for global evolution only if they inhibit the formation of the thin shell, which for the weak field scenario seems to be true qualitatively (Insertis & Rees 1991).

The entire question of the role played by cosmic rays and magnetic fields in *ellipticals* is an unanswered one and beyond the scope of this paper; thus we refer the reader to the many studies described above for discussions of their importance for wind generation in spiral discs. We re-iterate then, that in determining the timescale for gas ejection $t_{\rm GW}$, we have restricted ourselves to monitoring the influence of the gas thermal component only, with the caveat that substantial modification may be expected when a full accounting of cosmic ray and magnetic field energy is incorporated in future studies. For what follows, we provide the ejected masses at $t_{\rm GW}$, and note the mass of gas and metals which accumulates in the ISM in the quiescent post-wind phase (i.e. $t_{\rm GW} < t < t_{\rm G} \equiv 15.0$ Gyr). We then speculate that some fraction of this reservoir *may* be ejected by one or more of the following: ram pressure stripping by other galaxies or the ICM; smaller late-time bursts of wind inherent to the galaxy; cosmic ray-driven steady mass-loss; post-$t_{\rm GW}$ Type Ia SNe-driven steady mass-loss; etc.

As alluded to in Section 3, of the options presented herein, we feel Models $\rm B_3$ or $\rm B_3'$ are currently the best compromise for modeling $\varepsilon_{\rm th}$. Expansion of the shell is halted at the appropriate time (pressure equilibrium with the ambient ISM or contact with neighbouring shells), and radiative cooling proceeds until the temperature drops to $\sim 10^4$ K, beyond which the cooling becomes inefficient. Such a form for the thermal energy evolution leads to wind epochs which are two to three times earlier (see Table III) than those found by previous researchers who have used Model $\rm A_0$, and results in three to four times the mass of gas and metals being ejected from each elliptical. Putting this result into a full observational context is done in Gibson (1995), although two examples of the implications of adopting the CMB versus CC formalism are now discussed.

One obvious application of galactic wind models is in predicting the contribution of wind ejecta to the mass of gas and metals (in particular, Iron) observed in clusters of galaxies (e.g. Matteucci & Vettolani 1988; Renzini et al. 1993). We will not attempt a full analysis of the ICM abundance question here (see Gibson (1995) for more details), but will illustrate one example (Iron in the Virgo Cluster ICM) of the sensitivity of predicted ejected Iron masses to the assumed $\varepsilon_{\rm th}$ formalism. Our goal is to attempt to account for the observed Virgo ICM Iron mass (i.e. $(1.6-2.0)\times10^{10}$ M$_\odot$ - Arnaud 1994) within the confines of this SNe-driven galactic wind model.

Following the procedure outlined in Elbaz, Arnaud & Vangioni-Flam (1994), we first generated a series of power law fits relating a spheroid's present-day V-band luminosity to the mass of ejected Iron. The former was estimated using the fuel consumption theorem (FCT) of Renzini & Buzzoni (1986), coupled with the typical observed rest frame (B-V)$_0$ colours from Brocato et al. (1990). As in Padovani & Matteucci (1993), a multiplicative correction factor of 1.4 was applied to the FCT-determined $L_{\rm V}$. Two scenarios for the Iron ejecta were adopted. First, only the Iron ejected at $t_{\rm GW}$ was considered, as this "conservative" quantity is determined reasonably well by the wind model. Second, we estimate the maximum Iron available for deposition to the ICM (via one or more of the physical processes listed above) by adding in the mass of Iron which accumulates in the ISM reservoir during the post-$t_{\rm GW}$ regime. With these $M_{\rm Fe}^{\rm ej}$ versus $L_{\rm V}$ fits in hand, and the assumption of $L_{\rm V,E+S0}^{\rm Virgo}\equiv1.15\times10^{12}$ L$_\odot$ ($H_0\equiv50$ km/s/Mpc) (Elbaz, Arnaud & Vangioni-Flam 1994), we can predict the ICM Iron mass due to Virgo cluster ellipticals and lenticulars.

Table IV shows the predicted Virgo ICM Iron masses for both the "conservative" (i.e. $M_{\rm Fe}^{t_{\rm GW}}$) and "maximal" (i.e. $M_{\rm Fe}^{t_{\rm GW}+t_{\rm G}}$) models. Parallel predictions are made for the classic SNR thermal energy evolution Model A$_0$ and, for the first time, the newer CMB Models B$_3$ and B$'_3$. We have adopted both the Salpeter (1955) power law IMF of slope $x=1.35$, and the flatter $x=0.95$ slope preferred by Arimoto & Yoshii (1987).

Without resorting to the truncated IMF model of Elbaz, Arnaud & Vangioni-Flam (1994), the immediate conclusion to be drawn from Table IV is that the conservative $x=1.35$ IMF models are unable to account for the $(1.6-2.0)\times10^{10}$ M$_\odot$ of Iron observed. Model B$'_3$ is certainly an improvement upon Model A$_0$, which underproduces Iron by a factor of $\sim27$, but is still too low by a factor of $\sim6$. The "maximal" Model B$'_3$ with IMF $x=1.35$ is only a factor two lower than the observations. The flatter Arimoto & Yoshii (1987) IMFs coupled to the CMB formalism Models B$_3$ and B$'_3$ can easily reproduce the Virgo Iron abundance without resorting to any additional post-$t_{\rm GW}$ mass loss. It should be noted that Model A$_0$ with the flatter IMF can only recover the observations provided substantial late-time mass loss is invoked. The recent work of Elbaz, Arnaud & Vangioni-Flam (1994) does not support post-$t_{\rm GW}$ mass-loss at such a level.

The above discussion is by no means exhaustive, and is meant simply to be illustrative. Again, a more complete analysis is forthcoming (Gibson 1995).

Thus far we have ignored the effects of including a dark matter component distributed similarly to the luminous component. Specifically, we have assumed the ratio of dark matter mass to luminous mass $R$ in equation 7 to be zero. Taking $R=5\rightarrow10$, Matteucci (1992) rightly demonstrated that use of the Model A$_0$ formalism resulted in wind epochs of $\sim10$ Gyr for initial galactic luminous masses of $\gtrsim10^{11}$ M$_\odot$ (e.g. her Models C1 and C2). Recalling that active star formation does not cease until after the global wind (Larson 1974), Matteucci (1992) noted that based upon the "late" wind times, active star formation (i.e. $\psi = 1 \rightarrow 10$ $M_{\odot}$/yr) should be observed in low redshift (i.e. $z \lesssim 0.5$) ellipticals. Since this magnitude of star formation is not observed (e.g. Sandage 1986; Burkert & Hensler 1989) Matteucci (1992) concluded that the dark matter cannot be distributed like the luminous matter, but in a diffuse halo. While our aim is not to argue strongly against this conclusion, we would like to illustrate at least what the impact of changing from Model $A_0$ to Models $A_1$ or $B_3$ would have.

Following Bressan, Chiosi & Fagotto (1994), let us take $R = 5$ in equation 7. As expected from Matteucci's (1992) work, using the conventional Model $A_0$ for a $10^{12}$ $M_{\odot}$ model elliptical leads to $t_{\rm GW} \sim 13.5$ Gyr, and an implied star formation rate $\psi \sim 4$ $M_{\odot}$/yr at $z \sim 0$. Star formation at this level is what is observed currently in the local solar neighbourhood (i.e. late-type spiral disks), but not in typical nearby ellipticals. Using Model $A_1$ yields $t_{\rm GW} \sim 0.8$ Gyr, while Models $B_3$ and $B_3'$ give $t_{\rm GW} \lesssim 2$ Gyr. As we can see, using a SNR interior thermal energy evolution form which properly halts the expansion of the shell once pressure equilibrium between the interior and exterior ISM has been achieved (or once coming into contact with neighbouring shells), leads to wind epochs which occur early in the spheroid's lifetime, and hence active star formation at $z \lesssim 1$ is thereby avoided. Thus, based upon this chemical evolution argument alone, one should not dismiss outright the "dark matter distributed similarly to the luminous matter" scenario, as done by Matteucci (1992). Indeed, the recent work of Bertin et al. (1994) shows that there does exist many ellipticals whose photometric and kinematical data are consistent with the model in which the dark and luminous matter are distributed similarly, or even with an absence of dark matter altogether – five of the six ellipticals in their small sample fall into this category. This does not invalidate the diffuse dark halo model, as stressed by both Saglia, Bertin & Stiavelli (1992) and Bertin et al. (1994). We have illustrated simply that the dynamically dominant dark matter scenario can be accommodated within the framework of the classic SNe-driven wind model, provided one account properly for the late-time evolution of the SNRs internal thermal energies.

*6. Summary.* We have demonstrated the dangers in blindly adopting the SNR interior thermal energy evolutionary scenario in which each shell forever expands and cools in the pressure-driven snowplow phase. The most extreme example of this is seen when radiative cooling of the hot dilute interior is incorporated via the Cioffi, McKee & Bertschinger (1988) formalism; no galactic winds were found to develop in most of the model ellipticals unless the shell expansion was halted once coming into pressure equlibrium with the ambient ISM. Conversely, very early galactic wind epochs were predicted when the Cox (1972) and Chevalier (1974) form was coupled to this pressure equilibrium constraint due to the absence of interior cooling. We suggest the compromise scenario (our Models $B_3$ and $B_3'$) in which the late-time evolution of the thermal energy goes as $\varepsilon_{\rm th}(t) \propto t^{-1.04}$ (cooling and radiative cooling working together) until the shell either comes into pressure equilibrium with the ISM (Model $B_3$) or contact with neighbouring shells (Model $B_3'$). Cooling by radiative processes alone then leads to $\varepsilon_{\rm th}(t) \propto t^{-0.44}$ until the interior cools to $T \sim 10^4$ K, beyond which the efficiency of radiative cooling becomes negligibly small. Use of Model $B_3$, as opposed to Model $A_0$ which has been used predominantly in the past, leads to the onset of the galactic wind in approximately half the time. Consequently, the mass of gas and metals ejected is a factor of three to four times greater in the "newer" formalism.

When a dark matter component distributed similar to the luminous component is introduced, use of the Model $A_0$ formalism leads to late wind epochs (e.g. $t_{\rm GW} \approx 5 \rightarrow 10$ Gyr) for typical mass-to-light ratios, implying active star formation in $z \lesssim 1$ ellipticals. Since star formation of this magnitude is not observed, indirect arguments have been made in the past that the dark matter in ellipticals must then be distributed in a diffuse halo. Applying Models $A_1$ or $B_3$ we find that the predicted wind times occur significantly earlier in the spheroid's evolution (e.g. $t_{\rm GW} \lesssim 4$ Gyr, or $z \gtrsim 1$), and thus this indirect argument *alone* is not sufficient evidence for a diffuse dark matter distribution in ellipticals.

A conservative model for the Iron abundance in the Virgo Cluster ICM is proposed whereby a flatter than Salpeter ($x = 1.35$) IMF of slope $x \approx 1$, coupled with the CMB $B_3$ SNR thermal energy evolution models, can easily explain the measured metal mass. The classic Model $A_0$ can be salvaged only if a substantial amount of post-$t_{\rm GW}$ mass loss is invoked. The quantity of mass loss necessary seems at odds with the recent work of Elbaz, Arnaud & Vangioni-Flam (1994).

Finally, it is important to stress that our goal herein was to highlight simply a potential weakness in the SNR interior thermal energy evolution scheme adopted in previous galactic wind models. Using the newer Cioffi, McKee & Bertschinger (1988) formalism in place of that of Cox (1972) and Chevalier (1974) is a step in the right direction, but one must recognise that there are still inherent weaknesses in the one-phase elliptical models employed: *First*, we must re-iterate that the galactic ISM is not particularly well-described by the one-zone model. Two, three, and four-phase ISMs with varying degrees of inhomogeneities (Ostriker & McKee 1988) are certainly closer to reality, but difficult to quantify in a simple analytic form (although see Ferrini & Poggianti (1993) for a recent attempt to do so). *Second*, one must also recognise that there could be other significant contributors to an elliptical's ISM thermal energy, besides SNe. Magnetic field and cosmic ray energies play a substantial role in driving galactic winds in spirals, as their magnitudes are comparable to the SNe thermal energy, but a direct extrapolation to ellipticals is more problematic and will require future re-examination. Bressan, Chiosi & Fagotto (1994) have speculated recently that mass loss from stellar winds, and *not* SNe, is effectively the only important factor in driving a galactic wind, although this conclusion has been questioned by Gibson (1994).

These inherent weaknesses of the single phase, classic galactic wind model do *not* imply that analytical modeling of this sort is without merit. Even in this "flawed" state, its successful deployment in explaining, for example, the mass-metallicity relationship in ellipticals (e.g. Larson 1974) and abundance ratios in the ICM of galaxy clusters (e.g. Matteucci & Gibson (1994), and references therein), demonstrates its usefulness.

The aim of this paper is only to draw attention to the fact that the simplest analytical galactic wind models presented in the literature, thus far, have adopted a potentially inappropriate formalism for the SNR thermal energy evolution. Again, we do not presume to imply that use of the Cioffi, McKee & Bertschinger (1988) form provides a full and complete description of the ISM gas thermal evolution (due to the absence of cosmic ray and magnetic field energies, as well as the enforced ISM homogeneity). Our hope is that by illustrating the sensitivity of model predictions to the adopted SNR thermal energy formalism, we can stimulate further work towards quantifying analytically the influence of the aforementioned neglected ISM energy sources.

We have now begun the process of coupling our chemical evolution galactic wind code **MEGaW** to a full spectrophotometric package, the results of which will be presented elsewhere (Gibson 1995).

*Brad K. Gibson*
*Department of Astrophysics*
*Keble Road*
*University of Oxford*
*Oxford, U.K.*
*OX1 3RH*

*Acknowledgements.* I would like to thank Cedric Lacey, Chris McKee, Roger Chevalier, and Francesca Matteucci for a number of enlightening discussions and correspondences. The helpful suggestions of the anonymous referee are likewise acknowledged.

# TABLE I
## CMB FORMALISM:
## POWER LAW TIME DEPENDENT COEFFICIENTS GOVERNING THE DYNAMICAL ($\eta$, where $R_s \propto t^\eta$) AND THERMAL ($\nu$, where $\varepsilon_{th} \propto t^\nu$) EVOLUTION OF THE SNRs FOR THE MODELS DESCRIBED IN SECTION 2.2.1.

| Model | $t < t_{PDS}$ | | $a_2 t_{sf} < t < t_{merge}$ | | $t_{merge} < t < t_{cool}$ | | $t > t_{cool}$ | |
|---|---|---|---|---|---|---|---|---|
| | $\eta$ | $\nu$ | $\eta$ | $\nu$ | $\eta$ | $\nu$ | $\eta$ | $\nu$ |
| $B_0$ | 0.40 | 0.00 | 0.30 | – 1.04 | 0.30 | – 1.04 | 0.30 | – 1.04 |
| $B_1$ | 0.40 | 0.00 | 0.30 | – 1.04 | 0.00 | – 0.00 | 0.00 | – 0.00 |
| $B_2$ | 0.40 | 0.00 | 0.30 | – 1.04 | 0.00 | – 0.44 | 0.00 | – 0.44 |
| $B_3$ | 0.40 | 0.00 | 0.30 | – 1.04 | 0.00 | – 0.44 | 0.00 | – 0.00 |

Note: $\eta$ and $\nu$ vary smoothly from $t = t_{PDS}$ to $t = a_2 t_{sf}$. e.g. Figure 1.

# TABLE II

## CC FORMALISM: POWER LAW TIME DEPENDENT COEFFICIENTS GOVERNING THE DYNAMICAL ($\eta$, where $R_s \propto t^\eta$) AND THERMAL ($\nu$, where $\varepsilon_{\rm th} \propto t^\nu$) EVOLUTION OF THE SNRs FOR THE MODELS DESCRIBED IN SECTION 2.2.2.

| Model | $t < t_c$ | | $t_c < t < t_{\rm merge}$ | | $t > t_{\rm merge}$ | |
|---|---|---|---|---|---|---|
| | $\eta$ | $\nu$ | $\eta$ | $\nu$ | $\eta$ | $\nu$ |
| $A_0$ | 0.40 | 0.00 | 0.31 | – 0.62 | 0.31 | – 0.62 |
| $A_1$ | 0.40 | 0.00 | 0.31 | – 0.62 | 0.00 | – 0.00 |

# TABLE III
## GALACTIC WIND EPOCHS $t_{\rm GW}$ (Gyr) GAS FRACTION $f_{\rm g}$, METALLICITY $Z$, AND MASS OF METALS EJECTED $M_Z$ ($\rm M_\odot$) AT $t_{\rm GW}$. 'A' MODELS USE CC FORMALISM 'B' MODELS USE CMB FORMALISM

| $M_{\rm g}(0) = 1.0 \times 10^9$ $\rm M_\odot$ | | | | |
|---|---|---|---|---|
| Model | $t_{\rm GW}$ | $f_{\rm g}(t_{\rm GW})$ | $Z(t_{\rm GW})$ | $M_Z(t_{\rm GW})$ |
| $\rm A_0$ | 0.182 | 0.068 | 0.069 | 4.74(6) |
| $\rm A_1$ | 0.033 | 0.565 | 0.012 | 6.83(6) |
| $\rm A_1'$ | 0.033 | 0.565 | 0.012 | 6.83(6) |
| $\rm B_0$ | 1.468 | 0.003 | 0.065 | 1.72(5) |
| $\rm B_1$ | 0.059 | 0.374 | 0.023 | 8.68(6) |
| $\rm B_2$ | 0.229 | 0.040 | 0.080 | 3.22(6) |
| $\rm B_3$ | 0.102 | 0.197 | 0.041 | 8.09(6) |
| $M_{\rm g}(0) = 5.0 \times 10^{10}$ $\rm M_\odot$ | | | | |
| Model | $t_{\rm GW}$ | $f_{\rm g}(t_{\rm GW})$ | $Z(t_{\rm GW})$ | $M_Z(t_{\rm GW})$ |
| $\rm A_0$ | 0.570 | 0.016 | 0.077 | 6.09(7) |
| $\rm A_1$ | 0.118 | 0.299 | 0.030 | 4.55(8) |
| $\rm A_1'$ | 0.104 | 0.342 | 0.027 | 4.55(8) |
| $\rm B_0$ | n/a | n/a | n/a | n/a |
| $\rm B_1$ | 0.189 | 0.157 | 0.049 | 3.81(8) |
| $\rm B_2$ | 0.615 | 0.014 | 0.074 | 5.15(7) |
| $\rm B_3$ | 0.305 | 0.062 | 0.071 | 2.20(8) |
| $M_{\rm g}(0) = 2.0 \times 10^{12}$ $\rm M_\odot$ | | | | |
| Model | $t_{\rm GW}$ | $f_{\rm g}(t_{\rm GW})$ | $Z(t_{\rm GW})$ | $M_Z(t_{\rm GW})$ |
| $\rm A_0$ | 2.189 | 0.005 | 0.068 | 6.47(8) |
| $\rm A_1$ | 0.374 | 0.103 | 0.060 | 1.23(10) |
| $\rm A_1'$ | 0.307 | 0.147 | 0.051 | 1.50(10) |
| $\rm B_0$ | n/a | n/a | n/a | n/a |
| $\rm B_1$ | 0.477 | 0.063 | 0.070 | 8.79(9) |
| $\rm B_2$ | 1.859 | 0.006 | 0.068 | 7.88(8) |
| $\rm B_3$ | 0.813 | 0.020 | 0.077 | 3.05(9) |

## TABLE IV

PREDICTED VIRGO CLUSTER ICM IRON MASS (in $M_\odot$) CONTRIBUTIONS FROM EITHER GLOBAL GAS EJECTION AT $t_{\rm GW}$ ALONE ($M_{\rm Fe}^{t_{\rm GW}}$) OR $t_{\rm GW}$-COMPONENT PLUS COMPLETE POST-$t_{\rm GW}$ EJECTION OF ISM ($M_{\rm Fe}^{t_{\rm GW}+t_{\rm G}}$).

| $\varepsilon_{\rm th}$ Model | IMF $x$ | $M_{\rm Fe}^{t_{\rm GW}}$ | $M_{\rm Fe}^{t_{\rm GW}+t_{\rm G}}$ |
|---|---|---|---|
| $A_0$ | 1.35 | $5.93 \times 10^8$ | $4.28 \times 10^9$ |
| $A_1$ | 0.95 | $3.66 \times 10^9$ | $2.64 \times 10^{10}$ |
| $B_3$ | 1.35 | $1.47 \times 10^9$ | $7.36 \times 10^9$ |
| $B_3$ | 0.95 | $2.02 \times 10^{10}$ | $7.10 \times 10^{10}$ |
| $B'_3$ | 1.35 | $2.52 \times 10^9$ | $8.40 \times 10^9$ |
| $B'_3$ | 0.95 | $2.50 \times 10^{10}$ | $7.93 \times 10^{10}$ |
| Observed[a] | | $(1.6 - 2.0) \times 10^{10}$ | |

[a] Arnaud (1994)

## FIGURE CAPTIONS

**Figure 1:** Evolution of the interior thermal energy $\varepsilon_{\rm th}$ of a supernova remnant, normalised to $10^{50}$ erg. The sample models shown represent a total initial blast energy of $10^{51}$ erg, expanding into an ambient interstellar medium of number density $n_0 = 0.55$ cm$^{-3}$, mass density $\rho_0 = 9.2 \times 10^{-25}$ g/cm$^3$, and metallicity $Z = 0.02$.